\documentclass[aoas,preprint]{imsart}\usepackage[]{graphicx}\usepackage[]{color}
\makeatletter
\def\maxwidth{ %
  \ifdim\Gin@nat@width>\linewidth
    \linewidth
  \else
    \Gin@nat@width
  \fi
}
\makeatother

\definecolor{fgcolor}{rgb}{0.345, 0.345, 0.345}

\usepackage{framed}
\makeatletter
 {\par\unskip\endMakeFramed%
 \at@end@of@kframe}
\makeatother

\definecolor{shadecolor}{rgb}{.97, .97, .97}
\definecolor{messagecolor}{rgb}{0, 0, 0}
\definecolor{warningcolor}{rgb}{1, 0, 1}
\definecolor{errorcolor}{rgb}{1, 0, 0}

\usepackage{alltt}

\RequirePackage[OT1]{fontenc}
\RequirePackage{amsthm,amsmath,amsfonts}
\RequirePackage{natbib}
\RequirePackage[colorlinks,citecolor=blue,urlcolor=blue]{hyperref}
\usepackage{bm}
\usepackage{subcaption}
\usepackage{graphicx}
\usepackage{multirow}
\usepackage{xr}

\startlocaldefs
\numberwithin{equation}{section}
\theoremstyle{plain}

\newcommand{\yti}{Y_{Ti}}
\newcommand{\yci}{Y_{Ci}}

\newcommand{\etat}{\eta_T}
\newcommand{\etati}{\eta_{Ti}}

\newcommand{\mti}{\bar{m}_{Ti}}
\newcommand{\byt}{\bm{Y_T}}
\newcommand{\byc}{\bm{Y_C}}
\newcommand{\bmt}{\bm{\bar{m}_T}}

\newcommand{\EE}{\mathbb{E}}

\newcommand\numberthis{\addtocounter{equation}{1}\tag{\theequation}}

\endlocaldefs

\IfFileExists{upquote.sty}{\usepackage{upquote}}{}
\begin{document}

\begin{frontmatter}
\title{The Role of Mastery Learning in an Intelligent Tutoring System: Principal
  Stratification on a Latent Variable}
\runtitle{Principal Stratification for Mastery Learning}

\begin{aug}
\author{\fnms{Adam C} \snm{Sales}\ead[label=e1]{asales@utexas.edu}},
\and
\author{\fnms{John F} \snm{Pane}\ead[label=e2]{jpane@rand.org}}

\runauthor{A.C. Sales et al.}

\affiliation{University of Texas College of Education and RAND Corporation}

\end{aug}

\begin{abstract}
Students in Algebra I classrooms typically learn at
different rates and struggle at different points in the curriculum---a
common challenge for math teachers.
Cognitive Tutor Algebra I (CTA1), educational computer program,  addresses such
student heterogeneity via what they term ``mastery learning,'' where students progress
from one section of the curriculum to the next by demonstrating appropriate
``mastery'' at each stage.
However, when students are unable to master a
section's skills even after trying many problems, they are automatically promoted
to the next section anyway.
Does promotion without mastery impair the program's effectiveness?

At least in certain domains, CTA1 was recently shown to
improve student learning on average in a randomized effectiveness study.
This paper uses student log data from that study in a
continuous principal stratification model
to estimate the relationship between students' potential mastery and the CTA1
treatment effect.
In contrast to extant principal stratification
applications, a student's propensity to master worked sections here is
never directly observed.
Consequently we embed an item-response model, which measures students' potential
mastery, within the larger principal stratification model.
We find that the tutor may, in fact, be \emph{more} effective for students
who are more frequently promoted (despite unsuccessfully completing sections of the material).
However, since these students are distinctive in their educational
strength (as well as in other respects), it remains unclear whether
this enhanced effectiveness can be directly attributed to aspects of the mastery learning program.
\end{abstract}

\end{frontmatter}

\section{Introduction}
Teaching a class full of students who vary widely in ability is one of
the toughest challenges teachers face.
Intelligent tutoring systems may help.
These are pieces of software that are designed
to act as tutors, teaching material to individual students working on
computers \citep{anderson1985intelligent}.
Typically, they measure students' relevant skill sets and present them with personalized problems or exercises.
The students' performance on these exercises determines what they work
on next, based on updated measurements of their skill profiles.
This process is referred to as ``mastery learning''; students learn by
mastering skills, and only then moving on to new material \citep{bloom1968learning,kulik1990effectiveness}.
The hope is that by personalizing learning,
intelligent tutors can help teachers handle academic diversity.

\citet{pane2014effectiveness} reported the results of large-scale effectiveness study of the Cognitive Tutor Algebra I (CTA1), a curriculum whose centerpiece is the Cognitive Tutor software.
In the second year of implementation, the study found a moderate positive effect of CTA1 on high school post-test scores.

While CTA1 is designed around mastery learning, it does not always work
that way in practice \citep{descriptivePaper}.
For instance, the CTA1 system sets a maximum number of problems
students may work in each section.
Occasionally, some students will be unable to master a
set of skills before reaching the maximum number of problems.
In those cases, rather than allow the student's ``wheel-spinning''
\citep[c.f.][]{wheelSpinning} to continue indefinitely, CTA1 ``promotes'' them
to the next section.
Does the CTA1 treatment effect suffer as a result?
Do students who are more frequently promoted tend to experience
smaller treatment effects?

Student mastery is only defined subsequent to treatment
assignment---students in the control condition do not use the software
and therefore have no mastery data.
Traditional causal inference models, such as analysis of covariance
and subgroup analysis, can estimate the heterogeneity of treatment
effects as a function of pre-treatment covariates but cannot accommodate variables that may themselves be a function of the treatment.
On the other hand, principal stratification
\citep{frangakis,page2012principal,feller2016compared,sales2016student}
is designed for precisely such a task.
A principal stratification analysis could estimate the variance of treatment effects as a
function of \emph{potential} mastery: how often a student \emph{would} master worked sections, \emph{if} assigned to treatment.
This variable is defined prior to treatment assignment for all students in the study, but only observed for treatment students.

Implicitly, this assumes that potential mastery is measured without
error in the treatment group, an untenable assumption in our case.
There are no error-free measurements of students' propensity to master
sections.
Further complicating matters, both the number of worked sections and which sections
students worked varied widely between students in the treatment
group.
The typical principal stratification approach, assuming intermediate
variables measured without error, may yield misleading or
uninterpretable results when applied to mastery learning in CTA1.

This paper addresses the problem using a novel approach,
combining principal stratification modeling with item response theory
\citep[IRT; e.g.][]{embretson2013item} and latent variable analysis.
Using an IRT model to measure student mastery potential as a latent
variable brings a number of advantages over more traditional
approaches.
In particular, model-based measurement can account for variation in both the number of sections students work, and which sections students work, in addition to measurement error and missing data in general.
Defining principal strata based on latent variables may dramatically broaden the set of questions principal stratification may answer.

The structure of the paper is as follows.
The following section describes the CTA1 program, the CTA1 effectiveness trial, and the dataset.
Next, Section \ref{sec:principalStratification} reviews and illustrates continuous Bayesian principal stratification.
Section \ref{sec:modelingMastery} introduces an IRT model for mastery
and shows that it solves a number of problems with conventional
approaches.
Section \ref{sec:incorporatingIRT} discusses incorporating a latent
variable into Bayesian principal stratification.
Section \ref{sec:themodel} specifies a model for the CTA1 dataset and discusses its
identification; its results are in Section \ref{sec:results} and model
checks are in Section \ref{sec:modelChecking}.
Section \ref{sec:discussion} concludes with critical discussions of the methodological advances and the meaning of the model results for intelligent tutoring systems.

\section{Background}
\subsection{The Cognitive Tutor}
CTA1 is one of a series of complete mathematics curricula developed by
Carnegie Learning, Inc., which include both textbook materials and an
automated computer-based Cognitive Tutor
\citep{anderson1995cognitive}.
The CTA1 software divides the algebra course into units and sections within units. These are organized into a standard progression based on mathematics standards; however, schools have the option to customize these to meet local standards or other constraints. Many schools in the study exercised this option, meaning that although the basic set of sections and units is the same across the study, the sequence students encounter them is not.

The essential material of each section is represented as a set of
fine-grained knowledge components, or skills, and the software is
continually evaluating student mastery of these skills through the use
of a detailed computational model of student thinking in
algebra. Students solve problems and the model evaluates each student
action---whether it is a correct or incorrect action on a path toward
solving the problem, or a request for the software to provide a
hint---and updates its assessment of the mastery of each skill. When
students are judged to have mastered each skill in a section, they are
automatically moved to the next section. In an exception to this
general approach, when students work the maximum number of problems in
a section without mastering its skills, they are deemed to be ``wheel
spinning.''
The software promotes wheel-spinning students to the next section,
despite their non-mastery.
The system also enables teachers to override the mastery-based advancement to move students into a different section.

\subsection{The CTA1 Effectiveness Trial}
In 2007, the RAND Corporation received a grant from the
U.S. Department of Education to evaluate the  effectiveness of CTA1,
when implemented without any extraordinary support, in a diverse set
of schools. The project conducted two parallel experiments, one in 74
middle schools and one in 73 high schools, from 52 school districts in
seven states. Participating schools include urban, suburban, and rural
public schools, and some Catholic Diocese parochial schools, in Texas,
Connecticut, New Jersey, Alabama,  Michigan, and Louisiana. Each
school participated for two years. Schools in each state participated
in both the middle school and high school arms of the study except
Alabama (middle school only). 
Nearly 18,700 high school students 
participated in the study. 

The study used a blocked cluster randomized design to assign
schools to study condition. Schools within each state were matched
into pairs or triples, and randomized within blocks in the spring prior to their first year of implementation. Schools randomized to the treatment group implemented the CTA1 curriculum and those assigned to the control group continued to use their existing algebra I curriculum. Nearly all sites used materials published by Prentice Hall, Glencoe, or McDougal Littell. Assignments to treatment or control groups continued for two academic years in each school.

The study administered an algebra readiness pretest and an algebra proficiency post-test from the CTB/McGraw-Hill Acuity series. The exams were scored using a three-parameter IRT model. 
%
%
In the high school study, models estimated 95\% confidence intervals
for the treatment effect of
 -0.10$\pm$0.2 standard deviations in first year and
0.22$\pm$0.2 in the second year.
In the middle school study, models estimated 
treatment effect confidence intervals of -0.03 $\pm$0.2 the first year
and 0.19$\pm$0.3 the second year.

\subsection{Data for Principal Stratification}\label{sec:data}

Since our goal was to better understand the CTA1
treatment effect, we focused our analysis on data from high school students in
the second year of the CTA1 trial, for whom the treatment effect was
most evident.

We merged data from two sources: covariate, treatment and outcome data gathered by RAND, and computerized log data gathered by Carnegie Learning.
Table \ref{tab:covariateBalance} describes the covariates we used, including
missingness information, control and treatment means, and standardized
differences \citep[c.f.][]{kalton1968standardization}.
We singly-imputed missing values\footnote{We chose single imputation,
  instead of multiple imputation, for the sake of simplicity. Only
  pre-treatment data was used in the imputation process, so each
  imputed covariate is itself a pre-treatment covariate, and causal
  inference conditional on the imputed covariates is valid. That said,
  statistical inference regarding the covariates themselves, as in
  Section \ref{sec:predMast}, likely understates
  uncertainty.} with the Random Forest routine implemented
by the \texttt{missForest} package in \texttt{R}
\citep{missForest,rcite},  which estimated the ``out of box'' imputation
errors also shown in Table \ref{tab:covariateBalance} as part of the random forest regression.



\begin{table}[ht]
\centering

\begin{tabular}{cccrccc}
  \hline
  \hline
 &\% Miss.& Imp. Err.&Levels& Ctl.& Trt.& Std. Diff.\\
  \hline
  \hline
\multirow{3}{*}{Ethnicity}&\multirow{3}{*}{8\%}
  &\multirow{3}{*}{0.23}& White/Asian & 47\% &52\% & 0.16\\
&&&Black/Multi &32\% &26\% & -0.14\\
 & &&Hispanic/Nat.Am. & 21\% &22\% & -0.03 \\
\hline
\multirow{2}{*}{Sex}&\multirow{2}{*}{4\%}&\multirow{2}{*}{0.35}& Female & 51\% &49\% & -0.04\\
  &&&Male &49\% &51\% & 0.04\\
\hline
\multirow{3}{*}{Sp. Ed.}&\multirow{3}{*}{1\%}&\multirow{3}{*}{0.11}&Typical &87\% &86\% & -0.00\\
&&&Spec. Ed & 8\% &8\% & -0.02 \\
&&&Gifted &5\% &6\% & 0.03\\
\hline
Pretest&18\%&0.20&   &-0.33& -0.36& -0.05\\
   \hline
&&&\multicolumn{4}{c}{Overall Covariate Balance: p=0.22}\\
\hline
\hline
\end{tabular}

\caption{Missingness information and balance for the covariates included in this study, from
  the CTA1 Effectiveness experiment, high
  school, year two. Imputation error is percent falsely classified for
  categorical variables (Race/Ethnicity, Sex, and Special Education)
  and standardized root mean squared error for Pretest, which is
  continuous. 
  Analysis done in \texttt{R} via \texttt{RItools} \citep{ritools}.}
\label{tab:covariateBalance}
\end{table}

Over the course of the effectiveness study, Carnegie Learning gathered
log data from student users, including mastery or promotion for each section each student encountered.
Ninety-five control students (3\% of the control group) appeared in the mastery dataset, presumably because they transferred from schools assigned to the control condition to treatment schools.
We assumed that treatment assignment did not impact students'
decisions to transfer schools, and analyzed these students as control
students, excluding their mastery data from the analysis.

Log data were missing for some students, either because the log files were not retrievable, or because of an imperfect ability to link log data to other student records.
Treatment schools with mastery data missing
for 90\% or more students were omitted from the analysis, along with
their entire randomization block.
Of the remaining 2390 students, 84\% had mastery
data; treatment of missing mastery data for treatment students is
discussed below, in Section \ref{sec:incorporatingIRT}.

Mastery data for sections that were not part of the
standard CTA1 Algebra I curriculum, sections worked by fewer than 100
students, and sections that were mastered in every case were omitted from the
dataset.
The structure of the statistical model, described in section
\ref{sec:themodel}, justifies omitting these sections; a
sensitivity check including all Algebra I sections and every school in the
dataset yielded similar results.

Finally, because students' characteristics and behavior were of
primary interest, we included only data from worked sections that
ended in either mastery or promotion, omitting cases in which the teacher
moved the student to a new section prior to completion.

All told, the main analysis included $n=$5308 students, 2390 of whom were assigned to the CTA1 condition and 2918 of whom were assigned to control.
The students were nested within 116 teachers, in 43 schools across five states.
The analysis includes mastery information from 86,677 worked sections, 82\% of which were mastered.

\section{Principal Stratification for the CTA1 Experiment}\label{sec:principalStratification}
In the CTA1 experiment, let
$Z_i\in\{0,1\}$ represent student $i$'s treatment assignment,
$i=1,\dots,N$.
Let $Y_i$ denote $i$'s post-test score, so the central aim of the
experiment was to estimate the average effect of $Z$ on $Y$.
Following \citet{neyman} and \citet{rubin} let $\yti$ and $\yci$ denote $i$'s
``potential'' post-test scores were $Z_i=1$ or $Z_i=0$,
respectively---that is, were
$i$ assigned to treatment or control.
This notation implicitly assumes the ``Stable Unit Treatment Value
Assumption,'' or SUTVA \citep{sutva}: that there was only one version of the
treatment, and (since treatment was assigned at the school level) that
one school's treatment assignment did not affect outcomes in other schools.
Then the observed test score $Y_i=Z_i\yti+(1-Z_i)\yci$.
For each subject $i$ let $\bm{x}_i$ denote a vector of pre-treatment covariates.

Let $\tau_i=\yti-\yci$, $i$'s treatment effect.
Without strong untestable assumptions, $\tau_i$ is unidentified, since
for each $i$, either $\yti$ or $\yci$ is unobserved.
However, average treatment effects $\EE[\tau]=\EE[Y_T]-\EE[Y_C]$ are identified.
Similarly, randomization and SUTVA allow analysts to estimate treatment
effects conditional on a variable $x$, say $\EE[\tau|x]$, so long as
$x$ was not itself affected by treatment assignment---for instance,
gender or pretest scores.

The same cannot be said for so-called ``intermediate variables'' that are
themselves affected by treatment assignment.
Take $\bar{m}$, the proportion of a student's worked
sections that he or she mastered: $\bar{m}_i=\sum_s
m_{is}/n^{sec}_{i}$, where $m_{is}=1$ if student $i$ mastered section
$s$ and is zero otherwise.
$n^{sec}_{i}$ is the number of sections student $i$ worked,
$n^{sec}_i=\sum_s w_{si}$, where $w_{si}$ is an
indicator which equals one if student $i$ works section $s$ until
either mastery or promotion and zero otherwise.
Since the CT software was unavailable for control students, $\bar{m}$
is only defined for treatment students.
On the other hand, (following \citealt{frangakis})
let $\bar{m}_{Ti}$ represent the proportion of sections that $i$ would
master if assigned to the treatment condition---a potential value.
Unlike $\bar{m}$, $\bar{m}_T$ is defined for all subjects prior to
randomization, but only observed for members of the treatment group.
For a control student $i$ with $Z_i=0$, $\bar{m}_{Ti}$ is a counterfactual, representing
what \emph{would} have happened had $i$ been assigned to treatment,
i.e. had $Z_i=1$, counterfactually.
Randomization
guarantees that $\bar{m}_T$ is balanced---independent of treatment
assignment (conditional on school and randomization block).
Students who would master more sections, if given the
opportunity, were no more or less likely to be
assigned to treatment than those who would master fewer.

That being the case we may define a ``principal effect''
\citep[c.f.][p. 23]{frangakis} as the
super-population average treatment
effect, conditional on $\bar{m}_T$:
\begin{equation}\label{eq:principalEffect}
\tau(m)\equiv \EE[Y_T-Y_C|\bar{m}_T=m].
\end{equation}
Since $\bar{m}_T$ is a continuous variable,
\citet{gilbertHudgens} refer to $\tau(\cdot)$
as a ``causal effect predictiveness curve'' but we will follow
\citet{jin2008principal} and refer to $\tau(\cdot)$ as a principal
effect, as in the more typical case of a categorical intermediate
variable.
(Typical principal stratification also requires conditioning on $\bar{m}_C$---the proportion of sections
students would master if assigned to control---but this quantity is is undefined and
irrelevant in our case, and may be dropped from the analysis.)

Potential mastery $\bar{m}_T$ is observed for treated students (for
whom $\bar{m}_T=\bar{m}$), but unobserved for control students.
That said, randomization ensures that 
the distribution of $\bar{m}_T$ conditional on pre-treatment
covariates $\bm{x}$ is is the same in both treatment groups:
$\bar{m}_T|\bm{x},Z=1 =_d \bar{m}_T|\bm{x}, Z=0$ \citep[see][Lemmas
1 \& 2]{feller2016compared}.
These facts allow for partial identification of principal effects.

Principal effects may be estimated via randomization inference
\citep{nolen} or non-parametrically bounded \citep{bounding}.
Most commonly, they are estimated with a Bayesian model
\citep[e.g.][]{li2015evaluating,mattei2013exploiting}.
\citet{jin2008principal} and \citet{schwartz2011bayesian} give a full
treatment of Bayesian principal stratification with a continuous
intermediate variable such as $\bar{m}$, which we summarize here.
Let $\bm{Z}$, $\byt$, $\byc$ and $\bmt$ denote vectors of
students' treatment assignments, potential outcomes, and potential
mastery proportions, and let $\bm{X}$ denote the covariate matrix formed by
stacking row-vectors $\bm{x}^T$.
Then randomization implies that $\bm{Z}$ is independent of $\bmt$,
$\byc$ and $\byt$, and hence is ignorable.
Then, under exchangeability, for a vector of parameters $\bm{\theta}$
with prior density $f(\bm{\theta})$,
we may write the joint
distribution of $\byc$, $\byt$, and $\bmt$ as:
\begin{multline}\label{eq:fullModel}
f(\byc,\byt,\bmt|\bm{X})\\=\int \displaystyle\prod_i
f(\yci,\yti|\mti,\bm{x}_i,\bm{\theta})f(\mti|\bm{x}_i,\bm{\theta})f(\bm{\theta}) d\bm{\theta}.
\end{multline}
This formulation allows for posterior inference via Markov Chain Monte
Carlo techniques, such as data augmentation and
Gibbs samplers \citep{gelman2014bayesian}.

\sloppy
The model $f(\mti|\bm{x}_i,\bm{\theta})$ relates $\bar{m}_T$ to covariates.
Though $\mti$ is only observed for treated subjects, randomization
ensures that the same model holds for both treatment groups, and hence
is identified.
The model
$f(\yci,\yti|\mti,\bm{x}_i\bm{\theta})$, relates potential outcomes to
$\bar{m}_T$ and covariates.
The model for treatment potential outcomes
$f(\yti|\mti,\bm{x}_i,\bm{\theta})$, is entirely a function of
observed variables, and is non-parametrically identified.
In contrast, the model for control potential outcomes
$f(\yci|\mti,\bm{x}_i,\bm{\theta})$ depends on unobserved $\mti$;
therefore, our analysis must rely on an assumed model.
In practice, we we will assume that the model for $\yci$ is drawn from
the same family as $\yti$, albeit with different parameters; see
\citealt{richardson2011transparent} for an in-depth treatment of
analogous models.
Of course, the fit of the $\yci$ model may be compared to observed values $\yci$ and $\bm{x}_i$.

With these models in place, posterior inference for parameters
$\bm{\theta}$ proceeds by
separating the two models into treatment and control observations:
\begin{align*}
f(\bm{\theta}|\bm{Y},\bm{Z},\bm{X},\bm{M})&\propto\\
f(\bm{\theta})&\displaystyle\prod_{i: Z_i=1}f(\yti,\mti|\bm{x}_i,\bm{\theta})\\
\times&\displaystyle\prod_{i:Z_i=0} \int
        f(\yci|\bm{x}_i,\bm{\theta},\mti)f(\mti|\bm{x}_i,\bm{\theta})d\mti.\numberthis\label{eq:posterior}
\end{align*}
Though $\bar{m}$ is unobserved for members of the control group, its
conditional distribution may be estimated, and the marginal
distribution of $Y_C$ may be recovered via integration.

Fitting principal stratification models is fraught with challenges; even when the
model is well specified, multimodality and other pathologies of the
likelihood function can bias standard estimation procedures
\citep{griffin2008application,feller2016principal}.
These results make clear that any model-based principal stratification analysis must
include rigorous model checking and verification.

\begin{figure}
\centering
\includegraphics[width=0.8\textwidth]{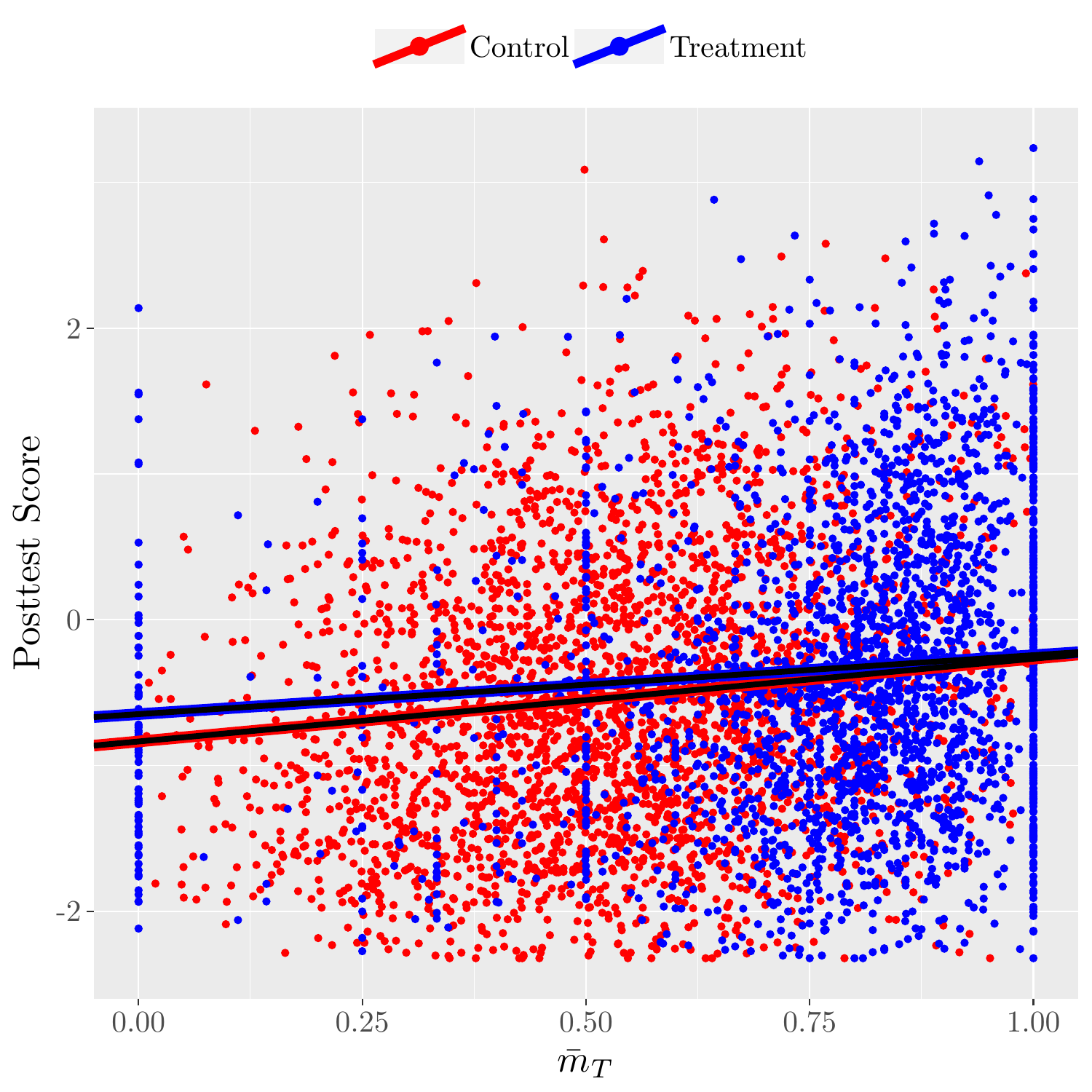}
\caption{A posterior draw from a principal stratification model
 stratifying on $\bar{m}_T$: observed posttest scores (in pooled
 standard deviation units) as
  a function of $\bar{m}_T$, along with regression lines.
$\bar{m}_T$ is observed for treated students
  group and imputed for control students.}
\label{fig:mBarModel}
\end{figure}

Figure \ref{fig:mBarModel} displays results from a principal
stratification model, with $Y_T$ and $Y_C$ and $\bar{m}_T$ modeled as linear in
covariates $\bm{x}$ with normally distributed errors clustered at the
teacher and school levels, and with
treatment effects linear in $\bar{m}_T$.
Details are available in an online supplement \citep{supplement}.
The x-axis of Figure \ref{fig:mBarModel} plots $\bar{m}_T$: for treated subjects, colored blue, the observed
value, and for control subjects, colored red, the 1000th MCMC draw.
The y-axis plots the observed posttest score $Y$.
The figure also shows the 1000th MCMC draws of regression lines from
the regressions of $Y_T$ and $Y_C$ on $\bar{m}_T$.
Though $\bar{m}_T$ appears positively correlated with achievement
in both treatment groups, the association is weaker in the treatment
group than in the control group.
This implies that students who would master a greater proportion
of worked sections, if assigned to treatment, tend to experience lower
treatment effects---CTA1 works best for students who tend to master
\emph{fewer} of the sections they work.
The full posterior distributions of the regression lines, estimated
via 4000 MCMC draws, are quite wide:
a 95\% credible interval for the difference between the lines' slopes was
 $[-0.3,0.2]$ pooled posttest standard deviations per one
 interquartile range (IQR) of $\bar{m}_T$.

\section{Modeling Mastery}\label{sec:modelingMastery}
\subsection{Problems with $\bar{m}$}\label{sec:problems-with-mbar}

\begin{figure}
\centering
\begin{subfigure}[b]{0.45\textwidth}
\includegraphics[width=\textwidth]{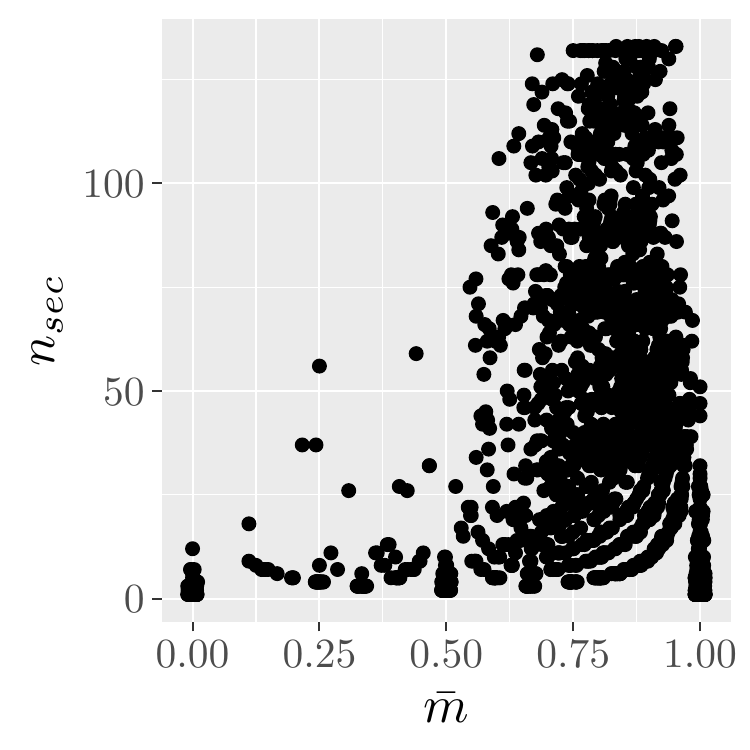}
\caption{ }
\label{fig:sampleSizeMbar}
\end{subfigure}
\begin{subfigure}[b]{0.45\textwidth}
\includegraphics[width=\textwidth]{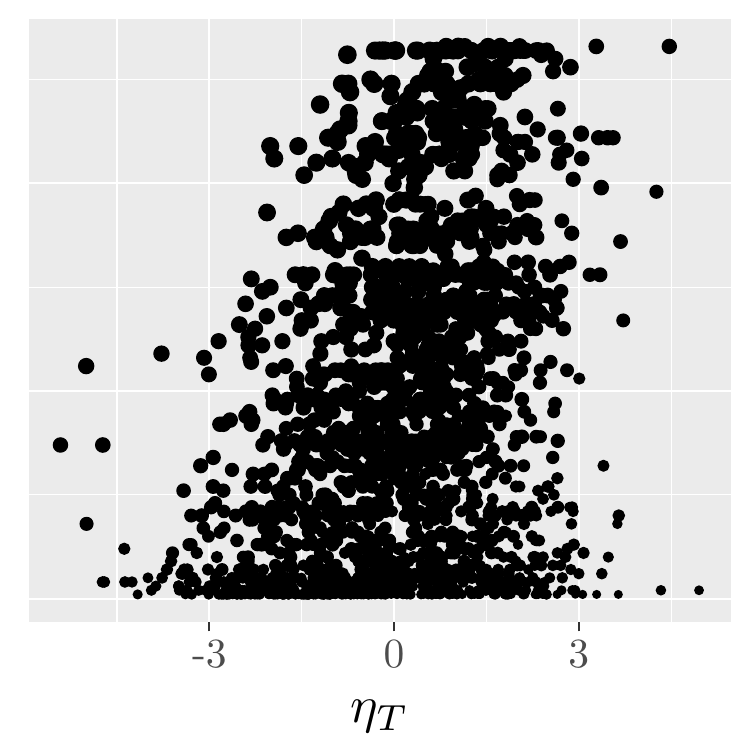}
\caption{ }
\label{fig:etaSampleSize}
\end{subfigure}
\caption{(a) Observed $\bar{m}$ as a function of $n^{sec}_i$. Overplotted
  points are jittered. (b) The 1,000th draw of $\etati$ for members of the treatment
  group as a function $n^{sec}$. $\etati$ is estimated from the model
  described by (\ref{eq:rasch1})--(\ref{eq:tauModel}).}
\end{figure}

The principal stratification model based on
$\bar{m}_T$, appears to be misspecified; Figure \ref{fig:mBarModel}
shows clear differences between the distribution of $\bar{m}_T$ in the
treatment group, observed as $\bar{m}$, and the distribution of imputed values for
$\bar{m}_T$ in the control group.
However, we shall see that even a well-specified model for
$\bar{m}_T$ would yield misleading results.

Figure \ref{fig:sampleSizeMbar} shows $\bar{m}$ as a function of
$n^{sec}$, the number of sections each student worked.
As one might expect, there is a strong correspondence: extreme low values
of $\bar{m}$ correspond almost exclusively to low values of $n^{sec}$.
This mechanism appears to drive the leftward skew of the $\bar{m}$
distribution, and complicates any interpretation of an estimated
function $\tau(m)$.
In particular, it is hard to disentangle the respective roles
$\bar{m}$ and $n^{sec}$ play in predicting treatment effects.

\begin{figure}
\centering
\begin{subfigure}[b]{0.45\textwidth}
\includegraphics[width=\textwidth]{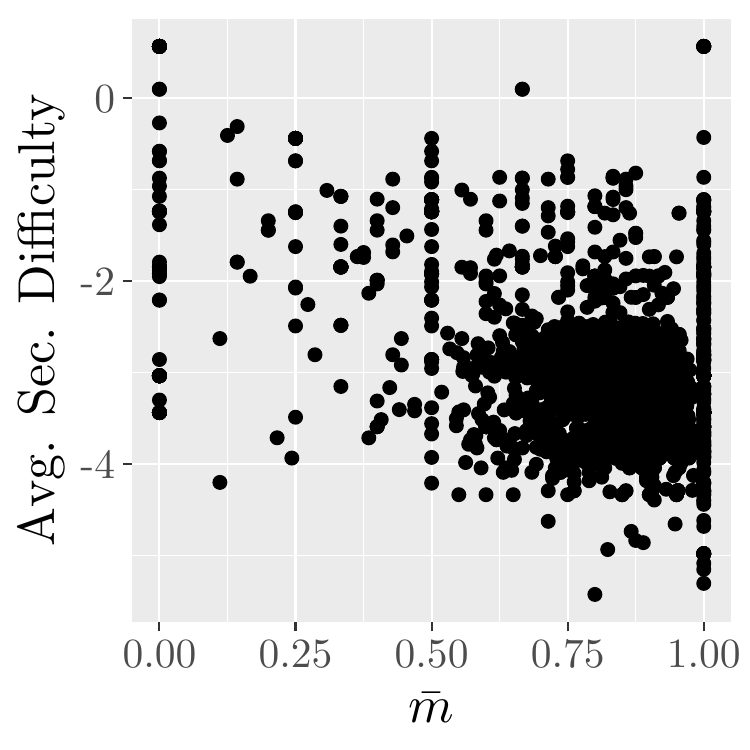}
\caption{ }
\label{fig:mbarDiff}
\end{subfigure}
\begin{subfigure}[b]{0.45\textwidth}
\includegraphics[width=\textwidth]{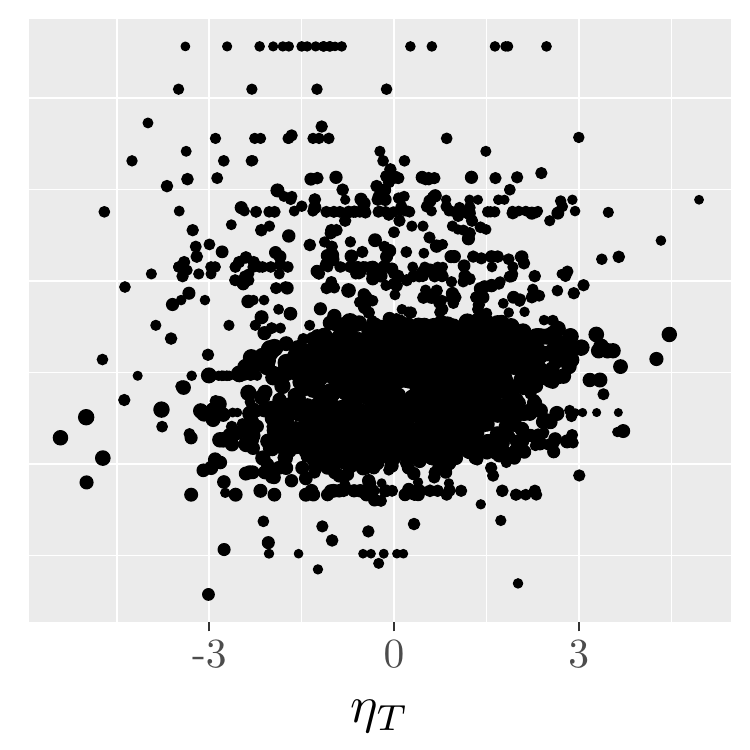}
\caption{ }
\label{fig:etaDiff}
\end{subfigure}
\caption{(a) Observed $\bar{m}$ for a student versus the average estimated
  difficulty of the sections he or she worked. Section difficulty was measured with
  the Rasch model (\ref{eq:rasch1})--(\ref{eq:rasch2}). The
  within-student averages used estimates from the 1000th posterior
  draw. (b) The 1000th draw of $\eta_T$ versus average section difficulty.}
\end{figure}

Students in the study vary not just in how many sections they attempt,
but also in which sections they work.
Figure \ref{fig:mbarDiff} plots each treated student's $\bar{m}$ as a
function of the average estimated difficulty of the sections he or
she worked (difficulty estimates are taken from the model we describe in
the next section).
A substantial amount of between-student variation in average section
difficulty is apparent in Figure \ref{fig:mbarDiff}---the difficulty
estimates are fixed effects from a logistic regression, so near the center of
their distribution a unit difference in section difficulty corresponds
to a difference of roughly 25\% in the probability of mastering a
section \citep[][p. 82]{gelmanHill}.
Unsurprisingly, students who work harder sections tend to master a
smaller proportion; the Spearman correlation
between $\bar{m}_T$ and average section difficulty is -0.30.
Therefore, $\bar{m}$ is only partially a measure of students' ability
to master worked sections---it also measures \emph{which} sections they worked.

\subsection{IRT Mastery Models}\label{sec:irt-mastery-models}
A better measurement of student mastery must account for variation in the number of
sections students worked, and their average difficulty.
This is closely related to one of the initial motivations for IRT: comparing students' scores across different
tests of the same material \citep{irtHandbook}.
In applying IRT terminology to CTA1 mastery data, the ``items'' are sections that students
work, and ``responses'' are binary indicators of mastery.
This statistical structure is analogous to educational and
psychological tests, the usual fodder for IRT models.
On the other hand, the substantive difference between mastery on sections
and responses to test questions requires careful attention.

Under the Rasch model \citep[e.g.][]{rasch}---perhaps the simplest
common IRT model---the probability that student $i$ masters worked section
$s$ is:
\begin{equation}\label{eq:rasch1}
Pr(m_{is}=1|w_{is}=1)=logit^{-1}(\etati-\delta_s)
\end{equation}
where $logit^{-1}(x)=\{1+exp(-x)\}^{-1}$ is the inverse logit
function.
The fixed ``difficulty'' parameter for section $s$, $\alpha_s$, in
this case reflects the difficulty of achieving mastery on section $s$.
The latent student ``ability'' $\etati$, modeled as a random
intercept, represents student $i$'s propensity to
master worked sections, since $Pr(m_{is}|w_{is}; \etati,\delta_s)$ increases with increasing
$\etati$.

Unlike in psychological testing, $\eta_T$ is not a measure
of student ability, knowledge, or achievement---though it may
correlate with these.
The Cognitive Tutor's principal aim is to
help students master algebra skills, so $\etati$
may be thought of as a measure of whether CTA1 \emph{works as
intended} for student $i$.
That is, mastery is CTA1's own criterion of success.
By definition, students who learn best from CTA1
are those who are more likely to
master the sections they work.

Model (\ref{eq:rasch1}) encodes a number of substantive
assumptions about how and when sections are mastered.
For instance, as suggested by an anonymous reviewer, it assumes that
the probability a student masters a section does not depend on which
other sections the student had previously worked.
Fortunately, students mostly adhered to the standard section order
imposed by Carnegie Learning, and almost always worked the sections in
an order that respected pre-requisite structure
\citep{descriptivePaper}.
Along similar lines, it also assumes that a student's propensity to master
a worked section remains constant over the course of the study.
This assumption would be violated if, for instance, students learned
over time how to better interact with the software, and were thus able
to master sections more reliably.
This, in turn, would induce a correlation between $\etati$ and
$n^{sec}_i$, so that the amount of student usage, and not just
mastery, impacts $\eta_T$.
Indeed, such a (rather slight) correlation appears to exist---for
instance, in Figure \ref{fig:etaSampleSize}---though it may be
due to other factors, such as student ability and motivation.
In supplemental analyses, when both the order in which each student
worked sections and $n^{sec}$ are
included in the model, the former appears to play little, if any,
role.
Finally, (\ref{eq:rasch1}) assumes that students' propensity to master sections can be
measured in one dimension---this would be violated if, for instance,
some students were more likely to master sections involving plotting,
but less likely to master sections involving solving equations for
unknowns, than their peers.
Section \ref{sec:modelChecking} examines the plausibility of this
assumption.

Modeling section mastery with an IRT model like (\ref{eq:rasch1})
addresses both of $\bar{m}_T$'s deficiencies.
Figure \ref{fig:etaSampleSize} plots $n^{sec}$ against a draw from the
posterior distribution of $\eta_T$ for treated subjects from the model
described below, in Section \ref{sec:themodel}.
Unlike $\bar{m}_T$, the distribution of $\eta_T$ do not skew left,
even for subjects with low $n^{sec}$.
The primary reason for this is partial pooling \citep{gelmanHill,rubin8schools,efronMorris}:
each individual's $\etati$ is estimated using both $i$'s data and data
from the rest of the sample.
Estimates of $\etati$ for students who worked few sections are shrunk
toward the overall mean $\EE \eta_T$, reducing the incidence of
outliers driven by noisy individual measurements.
Further, the posterior variance of $\etati$ conditional on
$\bm{m}_i=\{m_{is}\}_{w_{is}=1}$ depends on the number of problems
worked.
Fitting the measurement model (\ref{eq:rasch1}) simultaneously
with the rest of the causal model implicitly accounts for measurement
error in $eta_T$ \citep[e.g][Chapter 8]{carroll2006measurement}.

Figure \ref{fig:etaDiff} plots a posterior draw from  each student's
$\eta_T$ against the average estimated difficulty of the sections he
or she worked.
The negative relationship between difficulty and mastery, apparent in
Figure \ref{fig:mbarDiff}, is not present.
In fact, the posterior mean Spearman correlation between $\eta_T$ and average section
difficulty is positive 0.07, probably reflecting the fact that more capable
students are both more likely to master worked sections and more
likely to work on hard sections.
Unlike $\bar{m}_T$, variance in $\eta_T$ does not appear to be driven
by average section difficulty, but instead may reflect an underlying
student characteristic.

\section{Incorporating IRT into Principal Stratification}\label{sec:incorporatingIRT}

The $T$ subscript on the student parameter $\etati$ is not a common feature of IRT notation,
but is necessary due to $\etati$'s role in the causal principal
stratification model.
Fundamentally, it measures a baseline student
characteristic: what would be $i$'s propensity to master worked
sections \emph{were $i$ assigned to treatment}.
$\etati$, like $\bar{m}_{Ti}$, is a covariate, and, by definition,
is independent of randomized treatment assignment.
In other words, students with a greater propensity to master worked sections were
no more or less likely to be randomly assigned to the treatment condition than
students with a lower propensity for section mastery.

The parameter $\etati$ is well defined, if unobserved, for members of
the control group.
Therefore, its distribution, conditional on covariates, may be
extrapolated to the control group, in the
same way as $\bar{m}_T$.
In an ``explanatory'' Rasch model \citep[c.f.][]{de2013explanatory},
the student effects $\etati$ are modeled as a function of student
covariates, so that:
\begin{equation*}
f(m_{is}|w_{is},\bm{x}_i,\bm{\theta})=f(m_{is}|w_{is},\etati,\bm{\theta})f(\etati|\bm{x}_i,\bm{\theta})
\end{equation*}
where the density $f(m_{is}|w_{is},\etati,\bm{\theta})$ is as in
(\ref{eq:rasch1}), and the density $f(\etati|\bm{x}_i,\bm{\theta})$
plays a similar role to $f(\mti|\bm{x}_i,\bm{\theta})$ in
(\ref{eq:fullModel}).
Below we model $\etati$ as linear in covariates.

Unlike the model based on $\bar{m}$, (Section \ref{sec:principalStratification}), the
stratifying variable $\etati$ is unobserved for \emph{both} treated
and untreated subjects.
That said, the data available to estimate $\etati$ differ markedly
between the two groups.
For members of the treatment group, whose worked sections
$\bm{s}_i=\{s_{1i},\cdots,s_{n^{sec}_i,i}\}$ and mastery
$\bm{m}_i=\{m_{1i},\cdots,m_{n^{sec}_i,i}\}$ are observed, along with
covariates $\bm{x}_i$, the distribution of $\etati$ is a function of
all three (and other parameters $\bm{\theta}$):
$f(\etati|\bm{w}_i,\bm{m}_i,\bm{x}_i,\bm{\theta})\propto f(\bm{m}_i|\etati,\bm{w}_i,\bm{\theta})f(\etati|\bm{x}_i\bm{\theta})$.
On the other hand, members of the control group do not have data for
worked sections and mastery, so $\etati$ is only a function of
covariates and other parameters, $f(\etati|\bm{x}_i)$.
To estimate parameters $\bm{\theta}$, we have:
\begin{align*}
f(\bm{\theta}|\bm{Y},\bm{Z},&\bm{X},\{\bm{w}_i,\bm{m}_i\}_{i:Z_i=1})\propto\\
f(\bm{\theta})&\displaystyle\prod_{i: Z_i=1}\int f(\yti|\bm{x}_i,\bm{\theta},\etati)f(\bm{m}_{i}|\bm{w}_{i},\etati,\bm{\theta})f(\etati|\bm{x}_i,\bm{\theta})d\etati\\
\times&\displaystyle\prod_{i:Z_i=0} \int f(\yci|\bm{x}_i,\bm{\theta},\etati)f(\etati|\bm{x}_i,\bm{\theta})d\etati.\numberthis\label{eq:posteriorEta}
\end{align*}
To compute the posterior distribution of $\bm{\theta}$, it is
necessary to integrate over possible values of $\etati$ for all
subjects.

\sloppy
This structure also incorporates treated subjects with
missing mastery information: their contribution to the likelihood
integrates the density $f(\etati|\bm{x}_i,\bm{\theta})$ instead of the
density $f(\etati|\bm{x}_i,\bm{\theta},\bm{m}_i)$ as for other members
of the treatment group.
The model essentially multiply imputes $\etati$ for control students and
treatment students with missing mastery data.

\section{A Latent Principal Stratification Model for the Cognitive Tutor}\label{sec:themodel}
\subsection{Specifying the Model}

We modeled the probability that student $i$ achieved mastery on worked
section $s$ with the Rasch model (\ref{eq:rasch1}).
The model for latent mastery $\etati$ as a function of covariates
$\bm{x}_i$ was a normal regression:
\begin{equation}\label{eq:rasch2}
\etati|\left(\bm{x}_i,\bm{\theta}\right) \sim
\mathcal{N}\left(\bm{x_i}\bm{\beta}^M+\epsilon^{Mt}_{t[i]}+\epsilon^{Ms}_{s[i]},
\sigma^M\right)
\end{equation}
where $\bm{\beta}^U$ is a vector of coefficients.
Since students were nested within teachers, who were nested within
schools, we included normally-distributed school ($\epsilon_{Us}$) and
teacher ($\epsilon_{Ut}$) random intercepts.
The covariates in the model, $\bm{x}_i$, were detailed in Table
\ref{tab:covariateBalance}; preliminary model checking suggested
including a quadratic term for pretest, which was added as a column of
$\bm{x}_i$.
Although the entire principal stratification model is fit
simultaneously to both treatment groups, identification of the
parameters in (\ref{eq:rasch2}) comes primarily from subjects in the
treatment group for whom section mastery is observed.

We modeled students' post-test scores $Y$ as
conditionally normal:
\begin{multline}\label{eq:outcomeSubmodel}
 Y|\left(Z_i,\bm{x}_i,\bm{\theta},\etati\right) \sim \\ \mathcal{N}\left(
\beta^Y_{0b[i]}+\bm{x}_i^T\bm{\beta}^Y+a\etati+Z_i\tau(\etati)+\epsilon^{Yt}_{t[i]}+\epsilon^{Ys}_{s[i]},\sigma^Y_{Z[i]}\right)
\end{multline}
where $\beta^Y_{0b[i]}$ is a fixed effect for $i$'s randomization block, $\bm{\beta}^Y$ are the
covariate coefficients, and $\epsilon^{Yt}$, and
$\epsilon^{Ys}$ are normally-distributed teacher and school random
intercepts.
The residual variance $\sigma^Y$ varies with treatment assignment $Z$;
this captures measurement error in $Y$, treatment effect heterogeneity
that is not linearly related to
$\eta_T$, and other between-student variation in $Y$ that is not predicted by
the mean model.

Finally, we modeled treatment effects
$\tau(\etati)$ as linear:
\begin{equation}\label{eq:tauModel}
\EE[Y_{T}-Y_{C}|\eta_T]=\tau({\eta_T})=b_0+b_1\eta_T
\end{equation}
While more complex models for $\tau(\etati)$ are theoretically
possible (for instance, \citet{jin2008principal} uses a quadratic
model), the hypotheses that motivated this work predicted a monotonic
$\tau(\etati)$.
Additionally, more complex models for $\tau(\etati)$ tended to perform
poorly on the model checks described in \ref{sec:modelChecking}.

Covariates $\bm{X}$ were standardized prior to fitting.
Prior distributions for the block fixed effects $\beta^Y_b$ and covariate coefficients
$\bm{\beta^Y}$ and $\bm{\beta^M}$ were normal with mean zero and
standard deviation 2;
priors for treatment effects and the coefficient on $\eta_T$ were standard
normal.
The rest of the parameters received Stan's default uniform priors.
In all cases, we expected true parameter values to be much smaller in
magnitude than the prior standard deviation.

\subsection{Identifying and Fitting the Model}\label{sec:idAndFit}

We fit the model using the Stan software \citep{rstan} run from
\texttt{R} \citep{rcite}, simultaneously estimating all parts of the
model (\ref{eq:rasch1})--(\ref{eq:tauModel}).
We monitored convergence with traceplots and the Gelman-Rubin
statistic \citep{gelmanRubin}.

A secondary fitting exercise, based on multiple imputation \citep[e.g.][]{rubinLittle}, illustrates the factors that drive model
identification.
First, we extracted 1,000 MCMC posterior draws of $\etati$ from the
fitted model for all of the subjects in the dataset.
For treated subjects, these are similar to the standard ``ability''
scores from a Rasch model.
The difference is that these $\etati$ values incorporate data from
covariates $\bm{x}_i$, via (\ref{eq:rasch2}), and, more circuitously,
outcomes $Y$, since model (\ref{eq:rasch1})--(\ref{eq:rasch2}) were
fit simultaneously with (\ref{eq:outcomeSubmodel}).
Incorporating $Y$ into the model for $\eta_T$ is necessary if the
predicted $\eta_T$ values are to be used as imputations in a model for
$Y$ \citep[e.g.][]{sterne2009multiple}.
For subjects without usage data,
$\etati$ are random predictions from an explanatory Rasch
model.
Then, we fit 1,000 hierarchical linear models in \texttt{R}, using the
\texttt{lmer()} function from the \texttt{lme4} package \citep{lme4}.
In each regression $r$, we regressed outcomes on covariates $\bm{X}$ and a
treatment indicator interacted with the $r$th posterior draw for the
vector $\bm{\eta_T}$.
The distribution of estimates of the coefficient on the
treatment-$\eta_T$ interaction term was nearly identical to the
posterior distribution for $b_1$ described in the
next section---especially after scaling by the average variance of the
estimated coefficients.

This (perhaps didactic) exercise, we think, clarifies the inner
mechanisms of the full complex Bayesian model.
The explanatory Rasch model (\ref{eq:rasch1})--(\ref{eq:rasch2})
estimates $\etati$ for treated subjects and predicts it for control
subjects, and the outcome model (\ref{eq:outcomeSubmodel}) uses them
to estimate varying treatment effects.

\section{Results}\label{sec:results}

\subsection{Predicting Mastery}\label{sec:predMast}
\begin{figure}
\begin{subfigure}{0.8\textwidth}
\centering
\includegraphics[width=\textwidth]{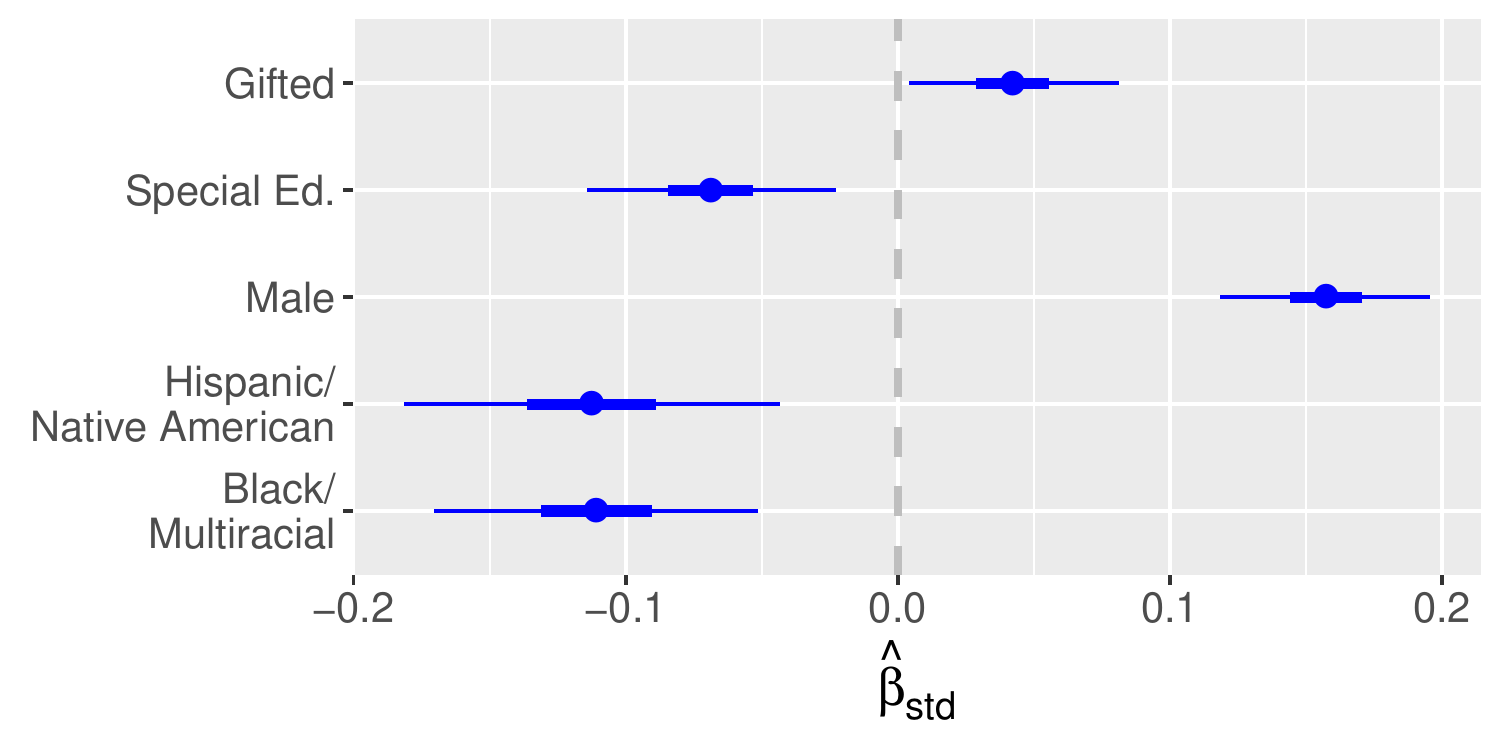}
\caption{}
\label{fig:usageCoef}
\end{subfigure}
\begin{subfigure}{0.8\textwidth}
\includegraphics[width=\textwidth]{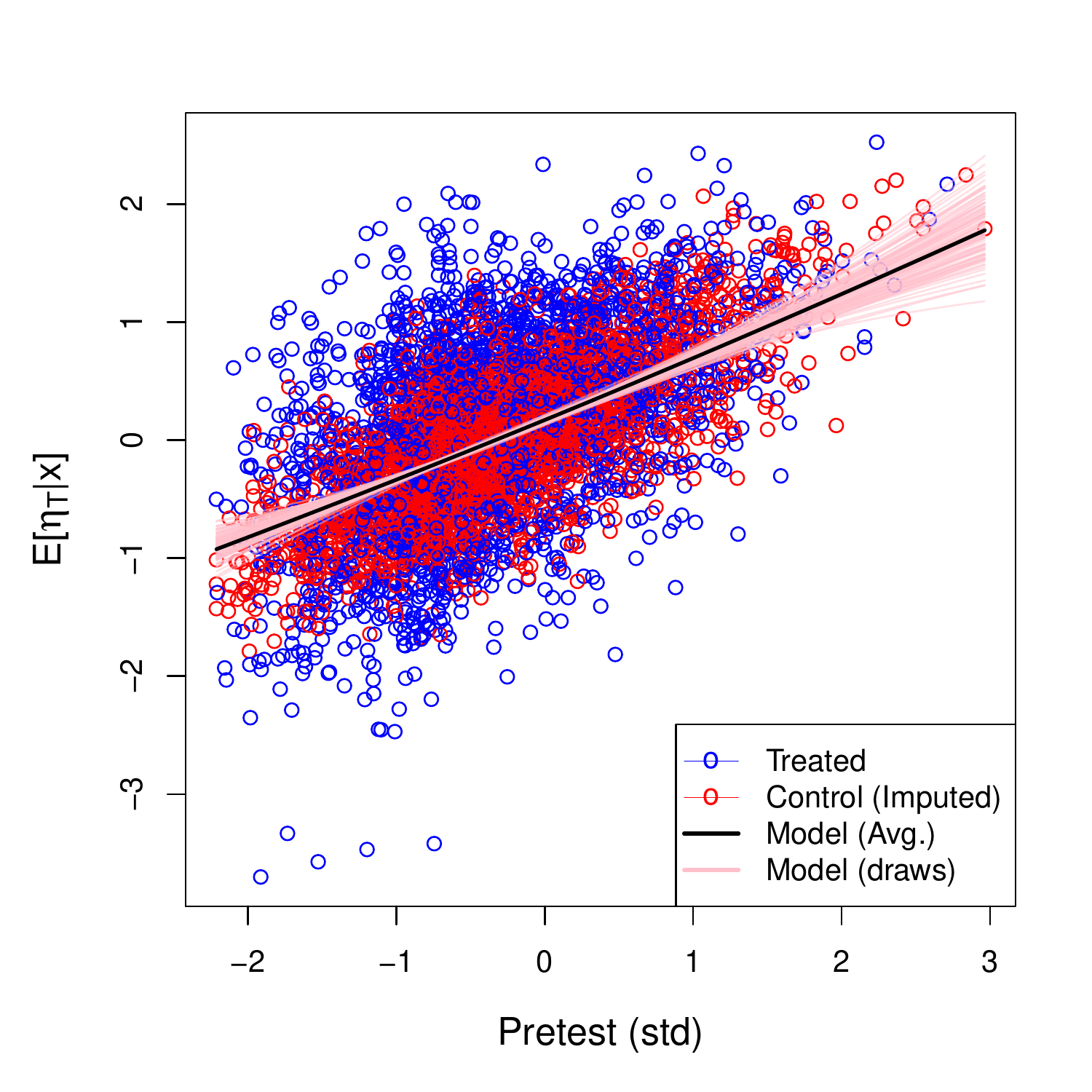}
\caption{}
\label{fig:usagePretest}
\end{subfigure}
\caption{Fitted model (\ref{eq:rasch2}), predicting $\eta_T$. (a)
  plots estimated coefficients on the categorical variables
  (race, with reference white/Asian, sex, with reference female, and
  special education, with reference typical) with 50\% and 95\% credible intervals. The coefficients
  represent differences in standard deviations of $\etati$. (b) plots
  estimated, or imputed, $\etati$ against pretest scores, with the
  posterior mean quadratic regression line and and 100 random
  posterior draws.}
\label{fig:usageResults}
\end{figure}

Which types of students are more, or less, likely to master worked
sections?
Figure \ref{fig:usageResults} displays the
estimated relationships between students' estimated $\etati$, or
$\EE[\etati|\bm{x}_i,\bm{m}_i,Y_i]$, and (singly-imputed) covariates $\bm{x}$.
Figure \ref{fig:usageCoef} displays the coefficients on five dummy
variables---two race categories, with White/Asian as the reference
category, and indicators for male, special education, and gifted
students.
The coefficients are standardized so that the units are in standard
deviations of $\etati$.
Figure \ref{fig:usagePretest} gives the relationship between pretest
scores and $\EE[\etati]$, by plotting $\EE[\etati]$ (standardized
similarly) by pretest, along with the estimated polynomial fit,
represented by the posterior mean and 100 random draws from the
posterior distribution of the regression line.

Apparently, students with higher pretest scores, white or Asian, male, and gifted
students are more likely to master worked sections. Black or
multiracial, Hispanic or Native American, special education students
and students with low pretest scores are less likely to master worked
sections.
On average, these variables, along with state indicators, explain
about 39\% of the variance in $\etati$.

Since the covariates in the model were singly-imputed using only
pre-treatment variables, these results must be interpreted with
caution.
An analysis with fully-observed variables or using multiple imputation
\citep{rubinLittle} may have generated different results.

\subsection{CTA1 Treatment Effects}\label{sec:ctai-treatment-effects}

\begin{figure}
\centering
\includegraphics[width=0.8\textwidth]{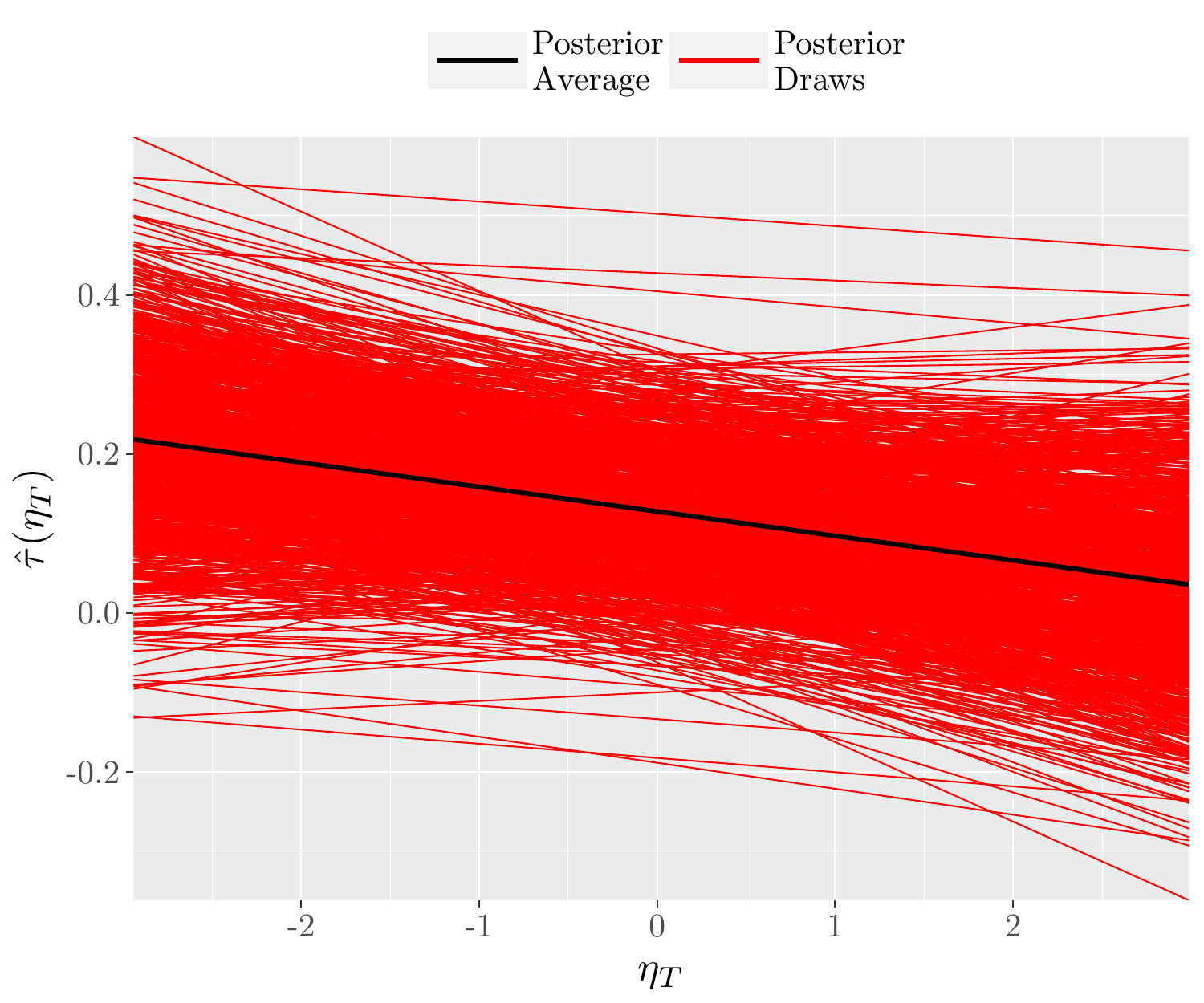}
\caption{The estimated treatment effect as a function of students'
  propensity to master a section, $\EE[Y_T-Y_C|\eta_T]$. Red lines are
  draws from the posterior distribution of the treatment effect
  function, and the black line is the mean of the posterior.}
\label{fig:trtEff}
\end{figure}

\begin{figure}
\includegraphics[width=0.8\textwidth]{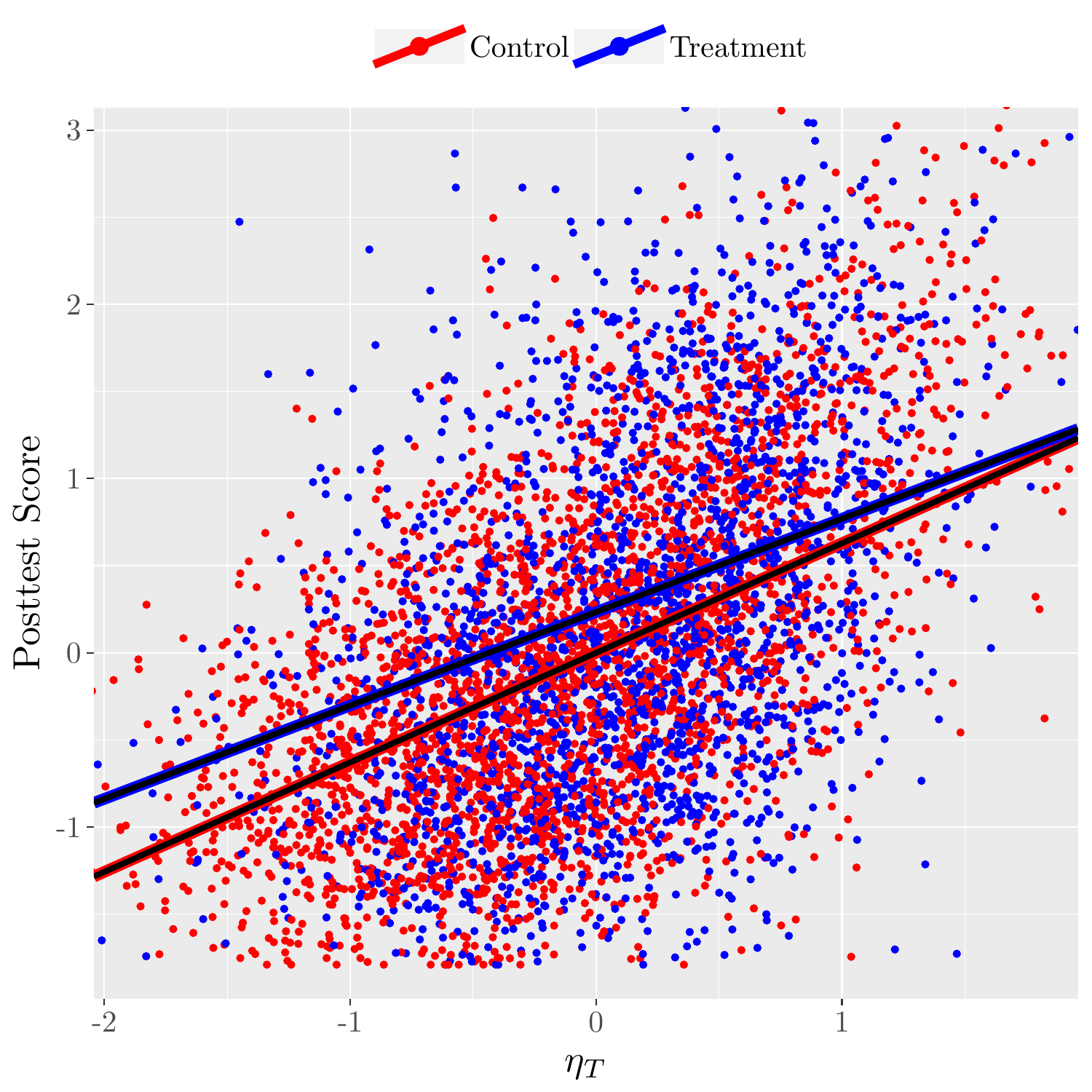}
\caption{One posterior draw from model
  (\ref{eq:rasch1})--(\ref{eq:tauModel}): observed post-test scores
  $Y$ (pooled standard deviations) as a
  function of estimated $\eta_T$ (in IQR units) in the control and treatment
  distributions, with regression lines.}
\label{fig:etaModel}
\end{figure}

Figure \ref{fig:trtEff} displays the posterior mean and posterior
draws for the estimated function $\tau(\etat)$.
These results suggest that, in fact, the treatment
effect decreased with increasing $\eta_T$.
Students who were more likely to master the sections they worked
experienced lower treatment effects.
Specifically, a difference of one IQR in $\eta_T$ was
associated with a reduction of 0.083 in the effect size
with a posterior standard deviation of 0.066.
In approximately 89\% of of the MCMC runs, the slope of $\tau(\etat)$
was negative; a central 95\% credible interval for the standardized
slope was [-0.212, 0.045].

Why might potential mastery be \emph{anti}correlated with treatment effects?
Figure \ref{fig:etaModel} plots observed outcomes $Y$ as a function of a
posterior draw of the vector $\bm{\eta_T}$.
(Fortunately, the misspecification apparent in Figure
\ref{fig:mBarModel}, based on $\bar{m}_T$ instead of $\eta_T$, does
not appear here.)
There is a positive relationship between $\etat$ and
$Y$ in both treatment groups---students who are more likely to master worked sections tend to
score higher on the post-test.
However, the slope between $Y$ and $\etat$ is slightly lower in the
treatment group than in the control group.
So as $\etat$ increases, the distance between $Y_T$ and $Y_C$---the
treatment effect---decreases.
These results suggest that CTA1 may be more effective for students who
would have scored lower on the post-test than for students who would
have scored higher.
This is unlikely to be the result of ceiling effects, since only one
student in the study correctly answered all posttest items.
The regression lines in Figure \ref{fig:etaModel} approach each other,
but do not cross---the model estimates a positive treatment effect for
the entire observed $\etat$ distribution.

It may seem surprising that our estimate for $\tau(\etat)$ 
would be more precise than our estimate for $\tau(\bar{m}_T)$ (Section
\ref{sec:principalStratification}), since $\bar{m}_T$ is partially
observed, whereas $\etat$ is completely unobserved.
Comparing posterior standard deviations,
the slope estimate for $\tau(\etat)$ was roughly twice as precise,
after standardizing the units.
The model misspecification evident in Figure \ref{fig:mBarModel} may be
partly to blame.
More importantly, both $\bar{m}_T$ and $\etat$ may be
thought of as measuring the same latent student quality---the
propensity to master a worked section.
By employing a more sophisticated and accurate measurement approach,
a model based on $\etat$ will often produce more precise estimates.

\section{Model Checking}\label{sec:modelChecking}
We checked the fit of model (\ref{eq:rasch1})--(\ref{eq:tauModel})
in several different ways,
using posterior predictive checks
\citep{rubin1984bayesianly,gelman1996posterior}, fitting a series of
alternative models, and fitting our main model to fake data, in
which the true parameters are known.

This discussion will focus on model checks aimed at two
central questions: first, does $\etat$ indeed measure potential
student mastery, and next, can our model successfully estimate real
treatment effect functions $\tau(\etat)$ without finding patterns
where none exists. A more complete list of model checks and their results is available in
the online supplement \citep{supplement}.

\subsection{Checking Measurement Validity}\label{sec:measVal}

Model \ref{eq:rasch1} assumes that students' propensity to master
worked sections is a unidimensional quantity.
To test this assumption, we conducted the posterior
predictive check described in \citet{levy2009posterior} using Yen's $Q_3$
discrepancy \citep{yen1993scaling}.
The median posterior predictive p-value
\citep[c.f.][]{zhu2011assessing} was 0.51, consistent with
approximate unidimensionality.
Additional details and results can be found in the online supplement \citep{supplement}.

Another concern is that $\eta_T$'s measurement of mastery is
confounded with overall student ability.
The CTA1 curriculum begins all students at the same place, regardless
of their initial ability.
Ideally, strong students quickly master more basic sections before
progressing to material they find more challenging in advanced
sections.
On the other hand, weaker students struggle with (and occasionally
fail to master) the first set of sections they encounter.
If this is the case, one would expect stronger students to achieve
mastery more often.
To allay this concern, we re-fit the principal stratification model using only data from
worked sections in which the student requested at least one hint.
This resulted in nearly identical results as our main analysis: a
difference of one IQR in $\eta_T$ was associated with a decrease of
0.081 in the treatment effect, with a standard
error of 0.065.

The online supplement \citep{supplement} discusses additional measurement validity checks,
including posterior predictive plots, and results from replacing the
Rasch model (\ref{eq:rasch1}) with a 2PL or 3PL model.
The results were nearly identical to those from our main model---for
instance, using a 3PL measurement model, we estimated that a
difference of one IQR in $\eta_T$ was associated with a decrease of
0.088 in the treatment effect size (SE: 0.068).

\subsection{Checking Estimation of $\tau(\etat)$}\label{sec:simulation}

We fit the model
(\ref{eq:rasch1})--(\ref{eq:tauModel}) to a series of
placebo datasets.
To create a placebo dataset, we dropped control schools, for which no
usage data is available.
We simulated a control group by duplicating outcome and covariate data
from the treatment group and relabeling the duplicate as the control
group.
The resulting dataset was comprised of a treatment group and a control
group, the former with usage data, but with exactly no treatment
effect (since the outcomes in the two groups were identical).
We created an additional three datasets by simulating treatment
effect functions $\tau(\etat)$ and adding them to the outcomes of the
``treated'' subjects: a randomly varying treatment effect uncorrelated
with $\etat$, and effects linear and quadratic in $\etat$.
Note that for the last dataset, in which effects are quadratic in
$\etat$, the linear model for $\tau(\etat)$ was misspecified.
For these models, we estimated $\etat$ by fitting
(\ref{eq:rasch1})--(\ref{eq:rasch2}) to usage data from the treatment
group.

\begin{figure}
\centering
\includegraphics[width=0.9\textwidth]{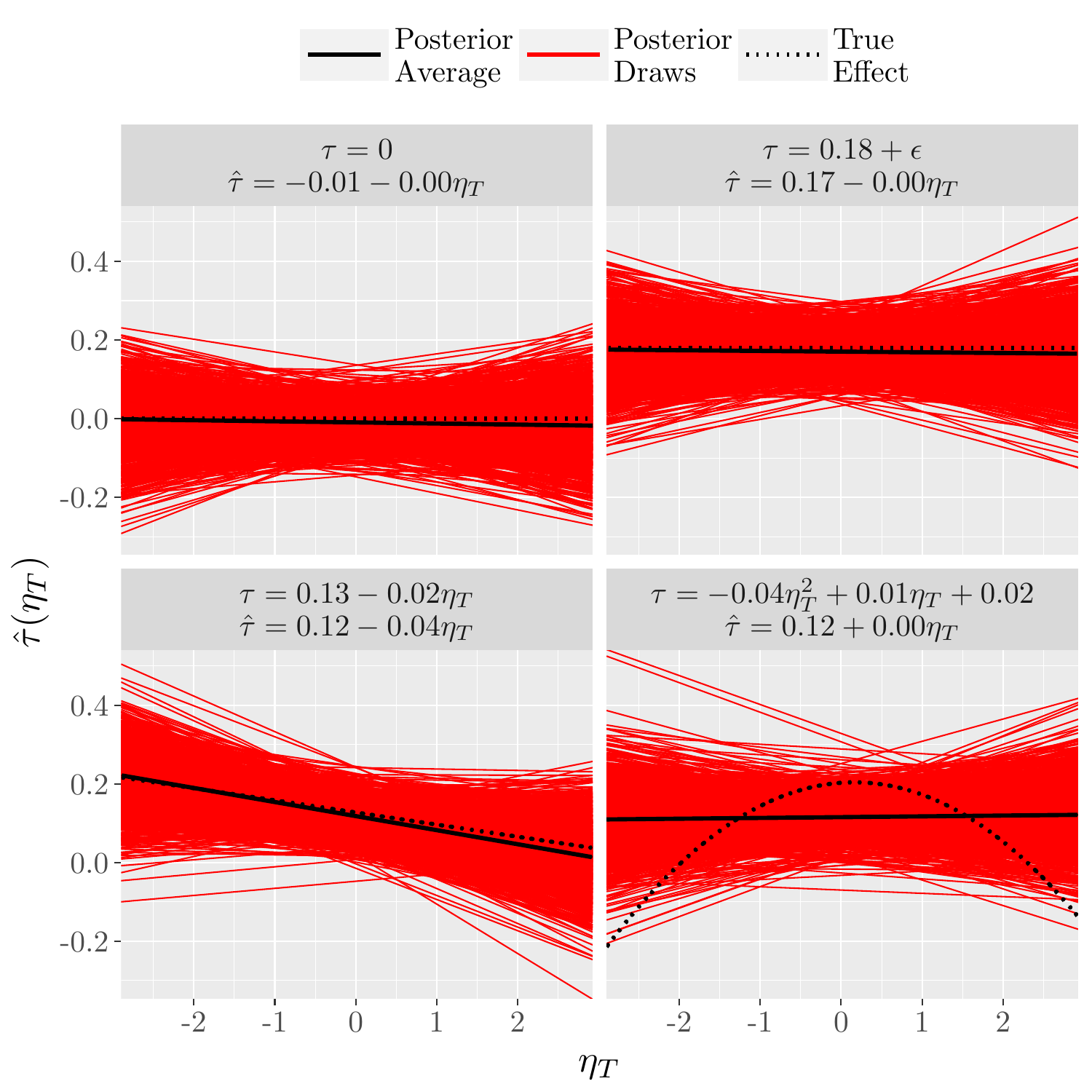}
\caption{Treatment effect estimates when model (\ref{eq:rasch1})--(\ref{eq:tauModel})
   was fit to
  datasets with simulated treatment effects, as described in Section
  \ref{sec:simulation}. In each figure, the solid black line represents the
  true treatment effect, the dotted line is the posterior mean of
  $b_1$, and the red lines are draws from the posterior distribution.
  The true and estimated (posterior mean) treatment effect functions
  are shown at the top of each panel. Clockwise from the upper left, there is no treatment effect,
  a random treatment effect uncorrelated with $\etat$,
  a treatment effect linear in $\etat$ and a treatment effect
  quadratic in $\etat$.}
\label{fig:sim}
\end{figure}

The results of fitting our model to these four datasets---one with no
treatment effect and three with simulated effects---are displayed in
Figure \ref{fig:sim}.
In the first three datasets, for which our treatment effect model was
well-specified, the model's estimates are in line with the truth.
In the final placebo dataset, in which the model was misspecified,
while the linear estimate of $\tau(\etat)$ fails to capture the true
pattern, it does lead to the correct conclusion of little or no
linear correlation between treatment effects and $\etat$.

\section{Discussion}\label{sec:discussion}

\subsection{The Role of Mastery in the Cognitive Tutor}

Were mastery learning the only driver of CTA1 effectiveness, we would
expect effectiveness to correlate with students' potential section
mastery.
In fact, the opposite seems to be the case---average effects appear to
decrease with students' mastery propensity (though they remain
positive throughout).

On the other hand, $\eta_T$, the latent parameter measuring mastery
propensity, positively correlates with students' pre- and post-test scores in both
treatment groups.
Students who were more likely to master worked sections were stronger at the
at the beginning of the year and knew more Algebra I at the end of the
year.
If the CTA1 effect were larger for lower performing students than for
stronger students we would expect to see a negative correlation
between $\eta_T$ and treatment effects.
Similarly, a wide range of pre-treatment student characteristics, both
measured and unmeasured, may explain the observed relationship between
$\eta_T$ and treatment effects.

Future work will extend this analysis to the middle
school arm of the study.
Middle school algebra students are not only younger than those in our
high-school sample, but are higher achieving as well, on average,
adding an interesting dimension to this analysis.
However, \citet{pane2014effectiveness} reported an imbalance in pretest scores of
treatment and control middle school students, suggesting that
assignment to the treatment condition may have influenced which students chose
to take algebra in middle school instead of waiting until high
school. This apparent selection into the treatment condition would have to be accounted for in a
principal stratification model, adding an additional modeling challenge beyond
those described here.

From a practical standpoint, these results are encouraging.
Fortunately, there is no evidence here that students' occasional failure to
master worked sections seriously impedes CTA1's effectiveness.
In fact, students who are more likely to wheel-spin may benefit even more from CTA1 than
their more successful peers.
Struggling students, who are less likely to achieve mastery, are also most in need of help.
The results here suggest that the Cognitive Tutor is not failing them.

\subsection{Latent Variables in a Potential Outcomes Framework}
In the course of modeling data from the CTA1 experiment, it became
necessary to introduce a latent variable into principal stratification
modeling.
We are unaware of this being done previously.
Latent variables are necessary here because directly observable
statistics ostensibly measuring student mastery---such as $\bar{m}$---
were woefully inadequate.
In particular, $\bar{m}$ does not account for which, or how many,
sections students attempted.
On the other hand, IRT provides a wealth of models
and a mature statistical theory for modeling student mastery potential.
Operationalizing students' potential mastery via the Rasch parameter
$\etat$ has clear advantages over the simpler approaches previously available.

That said, there may be some tension between latent variables and the
Rubin Causal Model, on which principal stratification is based.
For instance,
\citet[][p. 306]{IR} wrote:
\begin{quote}
Inferences across models with
different parametric structures can be compared directly because these inferences
are all driven by the posterior predictive distribution of the same causal
estimands defined by the potentially observable outcomes.
\end{quote}
One of the central arguments for the Rubin Causal Model is that its
target estimand is defined in a way that is independent of the model
used to estimate it.
In contrast, the definition of the parameter $\eta_T$ is inherently tied
to the Rasch model (\ref{eq:rasch1})--(\ref{eq:rasch2}).

But latent variables are themselves measurements.
The only difference between measurement via latent variables versus
via other measurement tools used in principal stratification is that
the measurement takes place within the principal stratification
model.
Perhaps the most common outcome in causal education research is test
scores, themselves typically calculated with an IRT
model---in other words, latent variables.
The models that give rise to the test scores are fit separately from
the causal model, giving them the appearance of objective
measurements.
Similarly, an analyst could fit model
(\ref{eq:rasch1})--(\ref{eq:rasch2}) to mastery and covariate data
without reference to outcomes, principal strata, or causal inference
at all.
Including the measurement model as a component of the larger causal
model is good statistical practice.

However, especially given the difficulty of fitting even much simpler
principal stratification models, an abundance of caution is in order.
Theory and guidance regarding when latent variable principal
stratification models will give accurate answers would be particularly
helpful.
The role of covariates in predicting latent variable values---and
hence imputing them for control subjects---is particularly pressing.

With the foundation set, latent variable principal stratification
can open many doors.
For instance, researchers may be able to use cluster analysis
techniques to summarize large numbers of intermediate
variables, and then examine treatment effect heterogeneity between
clusters.
Factor analysis may play a similar role in continuous principal
stratification.
Latent variable principal stratification has the potential to
facilitate more---and more nuanced---scientific discoveries.

\section*{Acknowledgments}
This material is based upon work supported by the National Science
Foundation under Grant Number 1420374. Any opinions, findings, and
conclusions or recommendations expressed in this material are those of
the author(s) and do not necessarily reflect the views of the National
Science Foundation. The authors wish to thank Brian Junker, Steve
Fancsali, Steve Ritter, two anonymous reviewers and the Associate
Editor for helpful input. 

\begin{supplement} 
\stitle{Supplement to: ``The Role of Mastery Learning in Intelligent Tutoring Systems: Principal
  Stratification on a Latent Variable''}
\slink[doi]{COMPLETED BY THE TYPESETTER}
\sdatatype{.pdf}
\sdescription{We provide modeling details, Stan code, and an extensive
  set of model goodness-of-fit and sensitivity analyses and plots.}
\end{supplement}

\bibliographystyle{plainnat}
\bibliography{ct}

\end{document}